\documentclass[review,number,sort&compress]{elsarticle}

\usepackage{amsmath,graphicx}
\usepackage{units}
\usepackage{rotating}

\begin{document}
\title{Maximum tunneling velocities in \\symmetric double well potentials}

\author[lsl,fubc]{J\"orn Manz}
\author[fubc]{Axel Schild}
\author[fubm]{Burkhard Schmidt}
\ead{burkhard.schmidt@fu-berlin.de, Phone: +49 30 838 75369, Fax: +49 30 838 75412}
\author[lsl]{Yonggang Yang}

\address[lsl]{State Key Laboratory of Quantum Optics and Quantum Optics Devices, 
Institute of Laser Spectroscopy, Shanxi University, 92, Wucheng Road, Taiyuan 030006, China}
\address[fubc]{Institut f\"ur Chemie und Biochemie, Freie Universit\"at Berlin, Takustr. 3, 14195 Berlin, Germany}
\address[fubm]{Institut f\"ur Mathematik, Freie Universit\"at Berlin, Arnimallee 6, 14195 Berlin, Germany}

\begin{abstract}
We consider coherent tunneling of one-dimensional model systems in non-cyclic or cyclic symmetric double well potentials.
Generic potentials are constructed which allow for analytical estimates of the quantum dynamics in the non-relativistic deep tunneling regime, in terms of the tunneling distance, barrier height and mass (or moment of inertia).
For cyclic systems, the results may be scaled to agree well with periodic potentials for which semi-analytical results in terms of Mathieu functions exist.  
Starting from a wavepacket which is initially localized in one of the potential wells, the subsequent periodic tunneling is associated with tunneling velocities. 
These velocities (or angular velocities) are evaluated as the ratio of the flux densities versus the probability densities.
The maximum velocities are found under the top of the barrier where they scale as the square root of the ratio of barrier height and mass (or moment of inertia), independent of the tunneling distance.
They are applied exemplarily to several prototypical molecular models of non-cyclic and cyclic tunneling, including  ammonia inversion, Cope rearrangement of semibullvalene, torsions of molecular fragments, and rotational tunneling in strong laser fields. 
Typical maximum velocities and angular velocities are in the order of a few km/s and from 10 to 100 THz for our non-cyclic and cyclic systems, respectively, much faster than time-averaged velocities.
Even for the more extreme case of an electron tunneling through a barrier of height of one Hartree, the velocity is only about one percent of the speed of light.
Estimates of the corresponding time scales for passing through {the narrow domain just} below the potential barrier are in the domain from 2 to 40 fs, much shorter than the tunneling times. 
\end{abstract}

\maketitle

\section{Introduction}

The time it takes for a particle to tunnel through a barrier is a topic that has attracted considerable interest already in the early days of quantum mechanics \cite{Hund:27a,Condon:31a,MacColl:32a}.
Closely related is the problem of the corresponding tunneling velocity.
Despite of its fundamental nature as well as its apparent simplicity, this question is still controversially discussed and a correct interpretation of tunneling is sometimes hampered by the absence of a unique definition of the tunneling time \cite{Landauer:94a,Winful:06a}. 
Some seemingly sensible {definitions, e.g.\ the group delay,} even lead to the predictions of particle velocities exceeding the speed of light \cite{Nimtz:11a}.
This effect of superluminality, known as the Hartman effect \cite{Hartman:62a}, was also claimed to be detectable in microwave experiments \cite{Enders:93a}. 
However, in more recent work it was suggested that this paradox does not violate relativity because the group delay time should not be interpreted as a transit time \cite{Davies:05a,Winful:06a}.

In this article, in contrast to the often used model of incoming free particles hitting a barrier and exiting freely \cite{Landauer:94a,Winful:06a}, and also in contrast with the hot topic of bound particles which dissociate or ionize through the time-dependent potential barrier induced by ultrashort intense laser fields (see e.g. Ref. \cite{Shafir:12a}), we study the case of coherent tunneling in a symmetric double well potential, i.~e. tunneling between bound states \cite{Hund:27a}. 
{Specifically}, we consider the deep tunneling regime of some model systems.
Here, the two delocalized wavefunctions of the lowest doublet of eigenstates with tunneling splitting $\Delta E$ can be superimposed with equal or opposite amplitudes thus forming two wavefunctions which are localized in one of the potential wells or the other, say either in the left or right one. 
These superposition states are not stationary. 
As a consequence, for example the left wavefunction will tunnel from the left potential well to the right one, and back, within tunneling time ${\tilde \tau} =h/\Delta {\tilde E}$. 
This well-known definition was derived by F. Hund, already in {1927 \cite{Hund:27a}. 
To the} best of our knowledge, however, the related topic of tunneling velocities ${\tilde v}$ during tunneling in the deep tunneling regime of {symmetric} double well potentials (assuming initial preparation as one of the superposition states which is localized e.g. in the left potential well) has not yet been considered in the literature. 
(The tilde notation {(like $\tilde{\tau}$)} refers to times, coordinates and velocities in terms of SI units; alternatively, for comprehensive derivations of the results we shall apply some convenient scalings of these variables, from SI to dimensionless units. The scaled variables will be written without tilde. 
The systems' parameters such as the mass $m$ of the tunneling particle, the barrier height $V_B$ of the double well potential and the positions $\pm x_0$ of the minima are also written without tilde.) 
The purpose of this paper is to derive a simple analytical expression which allows to estimate the maximum tunneling velocity, $\max {\tilde v}$, in this regime, in terms of few characteristic parameters such as $m$ and $V_B$. 
{Moreover, we shall compare the maximum tunneling velocity with the time-averaged one, $\operatorname{avg} \tilde{v} = 2 x_0 / (\tilde{\tau}/2)$}.

Our derivation of the maximum tunneling velocity, $\max {\tilde v}$, in systems with double well potentials will be restricted here to simple models of coherent tunneling along some coordinate $\tilde x$ which describes a (one-dimensional) path from the left potential well via the potential barrier $V_B$ to the right potential well. 
For convenience and for symmetry reasons, the position of the barrier will be defined as $\tilde x = 0$, and the minima of the left and right potential wells are located at $-x_0$ and $+x_0$, respectively. 
For this class of systems, the derivation will be rather general, that means we shall consider tunneling of systems with mass $m$ along non-cyclic Cartesian paths as well as systems with moments of inertia $I$ along cyclic (angular or torsional) paths.
For the corresponding velocities or angular velocities, we shall use the definition ${\tilde v} = {\tilde j}/{\tilde \rho}$ where $\tilde \rho$ and $\tilde j$ are the quantum mechanical probability densities and flux densities of the systems, depending on the coordinate $\tilde x$ and time $\tilde t$ which corresponds to analogous relations in classical mechanics and fluid dynamics {\cite{Winful:06a,Tannor:07}}.
Alternatively, this definition can also be obtained directly from the {time-dependent Schr\"odinger equation (TDSE)} and the polar representation of the wave function \cite{Bohm:52a,Wyatt:05a}. 
We shall show {that the} maximum value of $\tilde v$ is obtained just below the barrier, at ${\tilde x} = 0$. 
This implies a challenge because it is known that both the probability density ${\tilde \rho}(0)$ and the flux density ${\tilde j}(0)$ decrease exponentially when the barrier height $V_B$ increases \cite{Weiner:78a}. The limiting ratio of these two quantities is thus a priori unclear, and we shall particularly address the question whether $\tilde v$ is bounded or not. 
This question is in fact also motivated by the intriguing Hartmann effect, as outlined above \cite{Hartman:62a,Enders:93a,Nimtz:11a}. 
In order to answer this question, we shall take a risk by carrying out non-relativistic quantum dynamics simulations of the tunneling processes in terms of representative wavepackets which are obtained as solutions of the related {TDSE}. 
If the solutions of this approach would point to maximum tunneling velocities which approach the velocity of light, as reported for some cases of tunneling of free particles through potential barriers \cite{Enders:93a}, the present approach would have to be replaced by a relativistic one. 
Once we have determined the maximum tunneling velocity $\max {\tilde v}$, we shall also address the related question of the time ${\tilde t}_B = x_0/(5 \max {\tilde v})$ for passing through the ``most difficult part of the tunneling'', that means through the domain from ca. $-x_0/10$ to $+x_0/10$ just below the top of the barrier. 
{Moreover, we shall compare $\tilde{\tau}_{\rm B}$ with the tunneling time $\tilde{\tau}$. 
It will also be illuminating to compare the ratio $\tilde{t}_{\rm B}/\tilde{\tau}$ with the ratio of the time-averaged and maximum tunneling velocities, $\operatorname{avg} \tilde{v} / \max \tilde{v}$.}

The results which we shall derive below should be important for applications to many systems with symmetric double wells in chemistry and physics. 
For example, F. Hund in his fundamental paper \cite{Hund:27a} investigated tunneling from one enantiomer to the opposite one, with application to the torsional (cyclic) dynamics of H$_2$O$_2$. 
Below we shall consider complementary prototypical examples of molecules which may tunnel in cyclic symmetric double well potentials by torsional motions of two fragments about a connecting axis \cite{Manz:95a,Belz:13a,Grohmann:13a,Grohmann:07a,Barbatti:08a}. 
Alternatively, we shall also consider molecules which exhibit non-cyclic tunneling - the prototypical examples are tunneling of ammonia and semibullvalene along the coordinates which describe inversion \cite{Grohmann:13a,Letelier:97a} and Cope rearrangment \cite{Zhang:10a,Andrae:11a,Bredtmann:11a,Bredtmann:13a}, respectively.
In the context of this work, Refs. \cite{Grohmann:13a,Bredtmann:13a} are of special importance because they present not only the nuclear probability densities but also the first nuclear flux densities during tunneling in symmetric double well potentials, in the deep tunneling regime.
Finally, we point to the possible applications of rotational tunneling of molecules in external fields, induced by non-resonant interaction with laser fields through {anisotropic polarizability \cite{Zon:76a,Friedrich:91a,Friedrich:91b,Friedrich:95b,Friedrich:96,Stapelfeldt:03a}}. 
Very intense and short laser pulses are used to effectively align molecules, where the molecule-field interaction leading to laser--induced molecular alignment is given by a trigonometric potential energy function. This intimately connects to the general case of a pendulum in quantum mechanics \cite{Condon:28a,Pradhan:73a} for which the quantum dynamics of tunneling has recently been studied semi-analytically \cite{Leibscher:09a}.
Note that stationary pendular states can be expressed in terms of Mathieu functions \cite{Mathieu:68a}.
Although not analytically given, there is a substantial body of literature on their asymptotic properties \cite{Abramovitz:72a,Gradshteyn:80a,Meixner:54a}.

The article is organized as follows: 
In section \ref{sec:mosy}, we introduce a generic Hamiltonian which allows to consider tunneling of a non-cyclic as well as cyclic symmetric double well systems. 
The employed Hamiltonian depends on a single dimensionless action parameter $\beta$ which combines the effect of several system parameters, i.~e., the mass $m$ or moment of inertia $I$, the barrier height $V_B$ and width of the potential $x_0$.  
For this system, an expression for the potential is derived in section \ref{sec:copo} which is valid for sufficiently large values of $\beta$ and which is compared with a Mathieu model of pendular states. 
In section \ref{sec:tudy}, the tunneling dynamics of the lowest doublet in this potential is analyzed, and an analytic expression for the maximum tunneling velocity is found. 
Finally, in Sec. \ref{sec:discuss} we discuss these results, with reference to various applications. 
We shall also consider the consequences for the times ${\tilde t}_B$  which the systems need to pass through the domain just below the barrier, {together with the ratios $\tilde{t}_{\rm B}/\tilde{\tau}$ and $\operatorname{avg} \tilde{v} / \max \tilde{v}$.}. 
Sec. \ref{sec:discuss} also has some conclusions. 

\section{Model system and scaling properties}
\label{sec:mosy}

\subsection{Non-cyclic model}

Let us consider the case of a non-cyclic quantum system with mass $m$ tunneling along the coordinate 
$\tilde{x}$. The corresponding Hamiltonian is
\begin{equation}
\tilde{H} = -\frac{\hbar^2}{2m} \frac{\partial^2}{\partial \tilde{x}^2} + \tilde{V}(\tilde{x}).     
\label{eqn:htilde}
\end{equation}
The symmetric double well potential $\tilde{V}(\tilde{x})$ has its barrier 
centered at $\tilde{x} = 0$ with barrier height
\begin{equation}
 V_{\rm B} := \tilde{V}(0) - \tilde{V}(x_0),
\label{eqn:b}
\end{equation}
and with minima at $\tilde{x} = \pm x_0$.

The eigenfunctions $\tilde{\psi}_n(\tilde{x})$ and eigenenergies $\tilde{E}_n$ with quantum
numbers $n = 0,1,2,\dots$ are obtained as solutions of the time-independent
Schr\"odinger equation (TISE)
\begin{equation}
\tilde{H} \tilde{\psi}_n(\tilde{x}) = \tilde{E}_n \tilde{\psi}_n(\tilde{x}).     
\label{eqn:tisetilde}
\end{equation}

The model is thus characterized by three parameters, the mass $m$, the barrier
height $V_{\rm B}$ and width
parameter $x_0$. For the subsequent applications, it is convenient to introduce scaled, 
dimensionless variables $x = \tilde{x}/x_0$ and $E = \tilde{E}/V_{\rm B}$, i.e.\ the length 
and the energy are measured in terms of $x_0$ and $V_{\rm B}$.

Accordingly, we set $\tilde{V}(\tilde{x}) = \tilde{V}(x_0 x) = V_{\rm B} V(x)$, thus defining 
a scaled potential $V(x)$ with minima at $x = \pm 1$, and barrier height $V(0) - V(1) = 1$. 
In terms of scaled variables, the generic Hamiltonian $H = \tilde{H}/V_{\rm B}$ is
\begin{equation}
H = -\frac{\hbar^2}{2 m x_0^2 V_{\rm B}} \frac{\partial^2}{\partial x^2} + V(x) = -\frac{1}{2\beta^2} \frac{\partial^2}{\partial x^2} + V(x),
\label{eqn:h}
\end{equation}
where we have introduced the dimensionless parameter
\begin{equation}
\beta = \frac{\sqrt{m V_{\rm B}} x_0}{\hbar},
\label{eqn:betax}
\end{equation}
which is the action (in units of the reduced Planck's constant) also used in previous quantum and semiclassical treatment of tunneling, see, e.~g., Refs.~\cite{Miller:79a,Winful:06a}. 
Thus, the previous three parameters $m$, $V_{\rm B}$ and $x_0$ are replaced by just one parameter $\beta$ and the corresponding scalings of the energy and the length, $V_{\rm B}$ and $x_0$. 
This ensures that our results are transferable between systems with different values of their masses ($m$), barrier heights ($V_{\rm B}$), and widths ($x_0$), but with the same value of $\beta$. 
For a collection of $\beta$-values of systems relevant in chemical physics, see Table \ref{tab:tablee}, where $\beta$ is found to be in the range of 5 to 250. 
An electron which tunnels through a barrier of $V_{\rm B} =$ \unit[1]{E$_{\rm h}$} and $x_0 =$ \unit[1]{a$_0$}
yields $\beta = 1$. 
However, also in this case higher and/or broader barriers would yield larger values, similar to those in nuclear dynamics, as shall be considered later on.

The corresponding time-independent Schr\"odinger equation for the scaled
eigenfunctions $\psi_n(y)$ and eigenenergies $E_n$ is
\begin{equation}
H \psi_n(x) = E_n \psi_n(x),                                   
\label{eqn:tisescaled}
\end{equation}
with boundary conditions $\psi_n(x) \rightarrow 0$ for $x \rightarrow \pm\infty$,
normalization
\begin{equation}
\int_{-\infty}^{\infty} |\psi_n(x)|^2 dx = 1,                             
\label{eqn:normalization}
\end{equation}
and symmetries
\begin{eqnarray}
\psi_n(x) & =  \psi_n(-x) & \mbox{for} ~~~ n = 0,2,4,\dots \label{eqn:parityeven} \\
\psi_n(x) & = -\psi_n(-x) & \mbox{for} ~~~ n = 1,3,5,\dots \label{eqn:parityodd}
\end{eqnarray}
for even (\textit{gerade}, +) and odd (\textit{ungerade}, -) parities, respectively. The scaled
wave functions $\psi_n(x)$ are related to the original ones 
$\tilde{\psi}_n(\tilde{x})$ by
\begin{equation}
\psi_n(x) = \tilde{\psi}_n(\tilde{x}) \sqrt{x_0}. 
\label{eqn:psinscaled}
\end{equation}
An equivalent relation also holds for the time-dependent
wave functions $\tilde{\psi}(\tilde{x},\tilde{t})$ and $\psi(x,t)$, 
obtained as solutions of the scaled time-dependent Schr\"odinger equations (TDSE)
\begin{equation}
i \frac{\partial}{\partial t} \psi(x,t) = H  \psi(x,t),            
\label{eqn:tdsescaled}
\end{equation}
subject to proper boundary conditions (see below).
Here, we introduced a scaled time $t=\tilde{t} V_{\rm B} / \hbar$.
Hence, the velocity $v$ is measured in terms of $x_0 V_{\rm B}/\hbar$. 

\subsection{Cyclic model}

Alternatively, we shall also consider one-dimensional tunneling of a particle
with effective moment of inertia $I$ along an angle between $-\pi$ and $\pi$ 
in a symmetric double well potential with cyclic boundary 
condition. All equations for the non-cyclic case are also valid for the cyclic 
case, except that the mass $m$ is replaced by $I$, and
that the minimum positions are at $\pm \pi/2$. Thus, the dimensionless parameter $\beta$ is now defined as
\begin{equation}
\beta = \frac{\sqrt{I V_{\rm B}} \pi}{2\hbar}.
\label{eqn:betaphi}
\end{equation}
The time-dependent and time-independent wave functions obtained as solutions of 
the TISE (\ref{eqn:tisescaled}) and the TDSE (\ref{eqn:tdsescaled}), respectively, 
have to obey the cyclic boundary condition
\begin{equation}
\psi_n(2) = \psi_n(-2),                                          
\label{eqn:boundarycyclic}
\end{equation}
with normalization
\begin{equation}
\int_{-2}^{2} |\psi_n(x)|^2 dx = 1,                              
\label{eqn:normalizationcyclic}
\end{equation}
and symmetries (\ref{eqn:parityeven}), (\ref{eqn:parityodd}) for even (gerade, +) and odd
(ungerade, -) parities, respectively. 

\section{Potentials for analytical estimates of tunneling dynamics}
\label{sec:copo}

Using the generic model Hamiltonian (\ref{eqn:h}), which is parametrized by 
$\beta$ (\ref{eqn:betax}), (\ref{eqn:betaphi}), the next task is to construct an analytical 
model double well potential $V(x)$ which allows us to evaluate or estimate all the properties 
that are relevant for tunneling, ultimately the maximum (scaled) tunneling velocity 
$v$. Here, we focus on the deep tunneling regime, where 
\begin{equation}
\beta \gg 1.                                                   
\label{eqn:betalimit}
\end{equation}
This corresponds to rather high and/or broad potential barriers and/or large masses, 
or combinations thereof. Thus, the tunneling splitting
\begin{equation}
\Delta E = E_1 - E_0 
\label{eqn:tunnelsplitscaled}
\end{equation}
of the lowest doublet of levels becomes very small.

\subsection{Non-cyclic Gaussian model}

To begin, let us consider tunneling along the non-cyclic coordinate $x$.
We assume the ground-state wave function $\psi_0(x)$ to consist of two 
equivalent unimodal wave packets which do not overlap strongly,
\begin{equation}
 \psi_0(y) = N_0(\psi_{\rm l}(x) + \psi_{\rm r}(x)).
\label{eqn:psi0}
\end{equation}
The overall parity is $+$, eqn.\ (\ref{eqn:parityeven}). The functions 
$\psi_{\rm l/r}$ are localized close to the two equivalent potential minima 
at $x = \pm 1$ and have width $\Delta x$. Within the harmonic approximation 
of $V(x)$ at the left (``${\rm l}$'') and right (``${\rm r}$'') 
minima, the shapes of the two (normalized) wave packets approach Gaussians bell shapes, i.e.
\begin{equation}
 \psi_{\rm l/r} = \left(\frac{2}{\pi \Delta x^2} \right)^{1/4} \exp\left( -\frac{(x \pm 1)^2}{\Delta x^2}\right). 
 \label{eqn:psilr}
\end{equation}
Likewise, the first excited wave function consists approximately of the same
Gaussians, but with parity $-$, eqn.\ (\ref{eqn:parityodd}),
\begin{equation}
\psi_1(x) \approx N_1 (\psi_{\rm l}(x) - \psi_{\rm r}(x)).  
\label{eqn:psi1}
\end{equation}
Next, we seek a model potential $V(x)$ that yields the ground state wave function 
(\ref{eqn:psi0}) exactly. This potential is obtained by inverting the
TISE (\ref{eqn:tisescaled}),
\begin{eqnarray}
V(x) & = & \frac{1}{2 \beta^2 \psi_0(x)} \frac{\partial^2 \psi_0(x) }{\partial x^2} + E_0 \nonumber \\
     & = & \frac{ \left(\frac{2\left(x+1\right)^2}{\Delta x^2}-1\right) \psi_{\rm l} + \left(\frac{2\left(x-1\right)^2}{\Delta x^2}-1\right) \psi_{\rm r}}{\beta^2 \Delta x^2 (\psi_{\rm l} + \psi_{\rm r})} + E_0,
\label{eqn:www}
\end{eqnarray}
{see also \cite{Keung:88}.}

The relations between $\beta$, $\Delta x$ and $E_0$ are obtained by using the two known properties
of the scaled potential, i.e.\ $V(\pm 1) = 0$ and $V(0) = 1$, which yields
\begin{eqnarray}
\beta & = & \frac{\sqrt{2}}{\Delta x^2} \sqrt{\frac{\kappa - 3}{\kappa + 1}} \label{eqn:b1}\\
 E_0 & = & \frac{\Delta x^2}{2} + \frac{2\, \Delta x^2 - 4}{\kappa - 3} \label{eqn:E01}
\end{eqnarray}
with $\kappa = \exp(4/\Delta x^2)$ which yields real-valued solutions for $\Delta x \le 2/\sqrt{\log 3} \approx 1.9081$.

\begin{figure}[htbp]
\centering
\includegraphics[width=08cm]{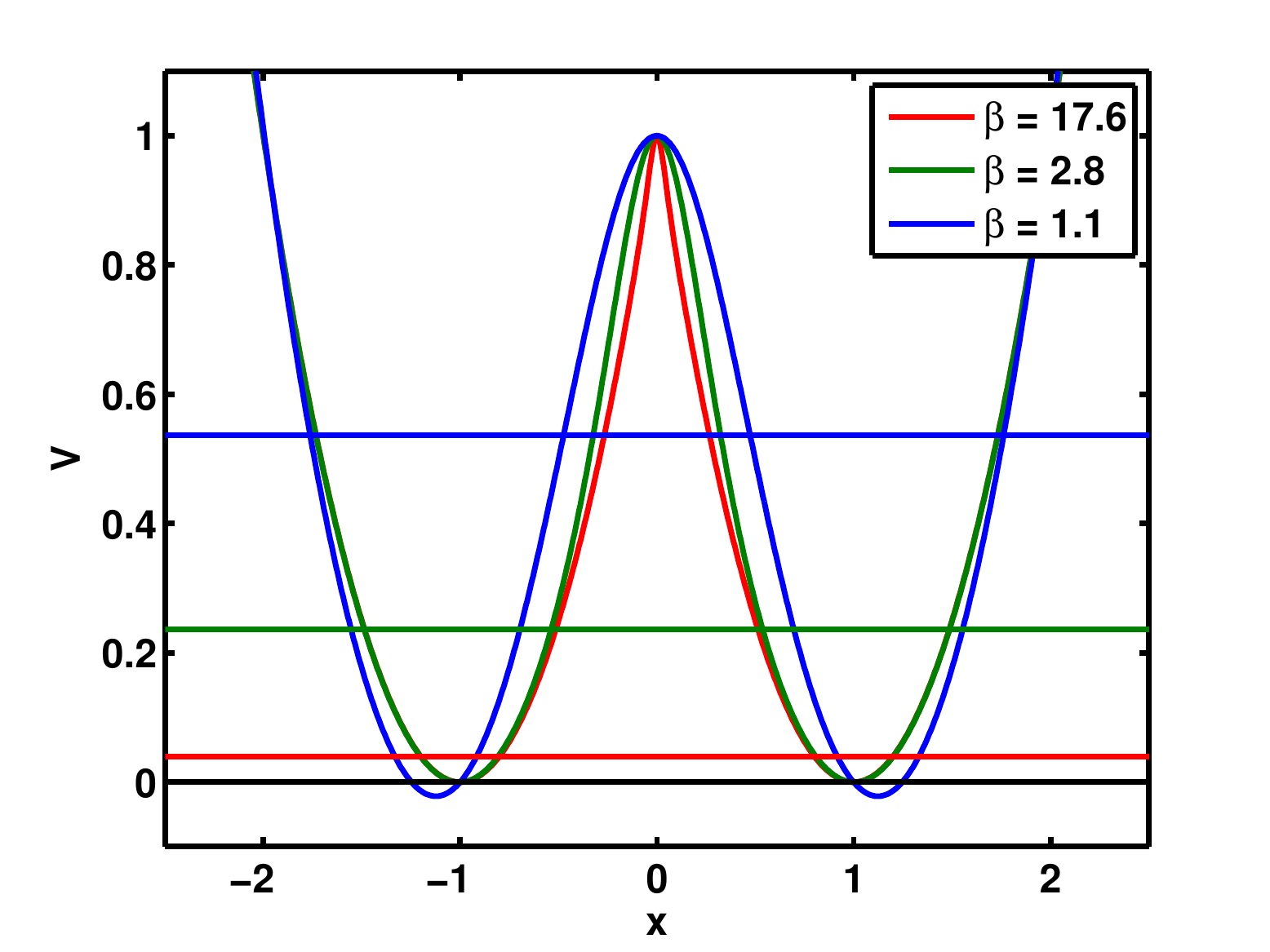}
\caption{Scaled symmetric double well potential $V(x)$, eqn.\ (\ref{eqn:www}), for the non-cyclic Gaussian 
model with dimensionless action parameters $\beta$ as specified in the legend, eqn. (\ref{eqn:betax}). The horizontal
colored lines indicate the respective zero point energies.}
\label{fig:pot_nc}
\end{figure}

Figure \ref{fig:pot_nc} shows the potential $V(x)$, eqn.\ (\ref{eqn:www}), 
for the parameters $\beta = 1.1, 2.8$ and $17.6$ (corresponding to $\Delta x 
= 1.10, 0.69$ and $0.28$, respectively), together with the eigenvalues $E_0$. 
While all curves fulfill the above 
conditions, we note that for small $\beta$ the two unimodal wave packets (\ref{eqn:psi0}) start to overlap significantly, so that the minima are slightly shifted 
outwards. Thus, from now on we restrict ourselves to the case of well-localized
wave packets at the minima of the potential, i.e.\ small $\Delta x$ and consequently
large $\beta$, equivalent to the deep tunneling condition of eqn.\ (\ref{eqn:betalimit}). 
In this limit, the minima approach indeed $x = \pm 1$, and
eqns.\ (\ref{eqn:b1}), (\ref{eqn:E01}) simplify to 
\begin{eqnarray}
\beta & = & \frac{\sqrt{2}}{\Delta x^2} \\
 E_0 & = & \frac{\Delta x^2}{2},
\end{eqnarray}
which will be used throughout the remainder of this article.
At the same time, the two normalization constants $N_0$ and $N_1$
in eqns.\ (\ref{eqn:psi0}) and (\ref{eqn:psi1}) approach their limiting
values, $1/\sqrt{2}$.

The tunneling splitting (\ref{eqn:tunnelsplitscaled})
can be approximated for small tunneling splitting, as derived in \cite{Weiner:78a}, by
\begin{eqnarray}
\Delta E & \approx -\frac{1}{\beta^2} \psi_0(0) \left.\frac{\partial \psi_1(x)}{\partial x}\right|_{x=0} \nonumber \\
& = 2^{5/4} \sqrt{\frac{2}{\beta \pi}} \exp(-\sqrt{2}\beta).                  
\label{eqn:deltaeweiner}
\end{eqnarray}
The dependence of $\Delta E$
on $\beta$ is shown in the upper panel of Figure \ref{fig:velocities}. Within the considered
range of $\beta$, the tunnel splitting $\Delta E$ decreases by many orders of magnitude.

\subsection{Cyclic Gaussian model}

For cyclic double well potentials, we use the same expression (\ref{eqn:www}) for the 
potential $V(x)$ in the domain $-1 \le x \le 1$, and repeat it periodically.
This modifies the non-cyclic potential $V(x)$ in the 
domain of the second barrier close to $x = \pm 2$, without any significant 
effects at the potential minima, in the limit (\ref{eqn:betalimit}). 
For convenience, the minima at $x = \pm 1$ are called the left and right potential minima, 
and also for the case of the cyclic double well potential we use (\ref{eqn:psi0}) and (\ref{eqn:psi1}) 
as approximations to the exact wave functions $\psi_0(x)$ and $\psi_1(x)$. 

\begin{figure}[htbp]
\centering
\includegraphics[width=08cm]{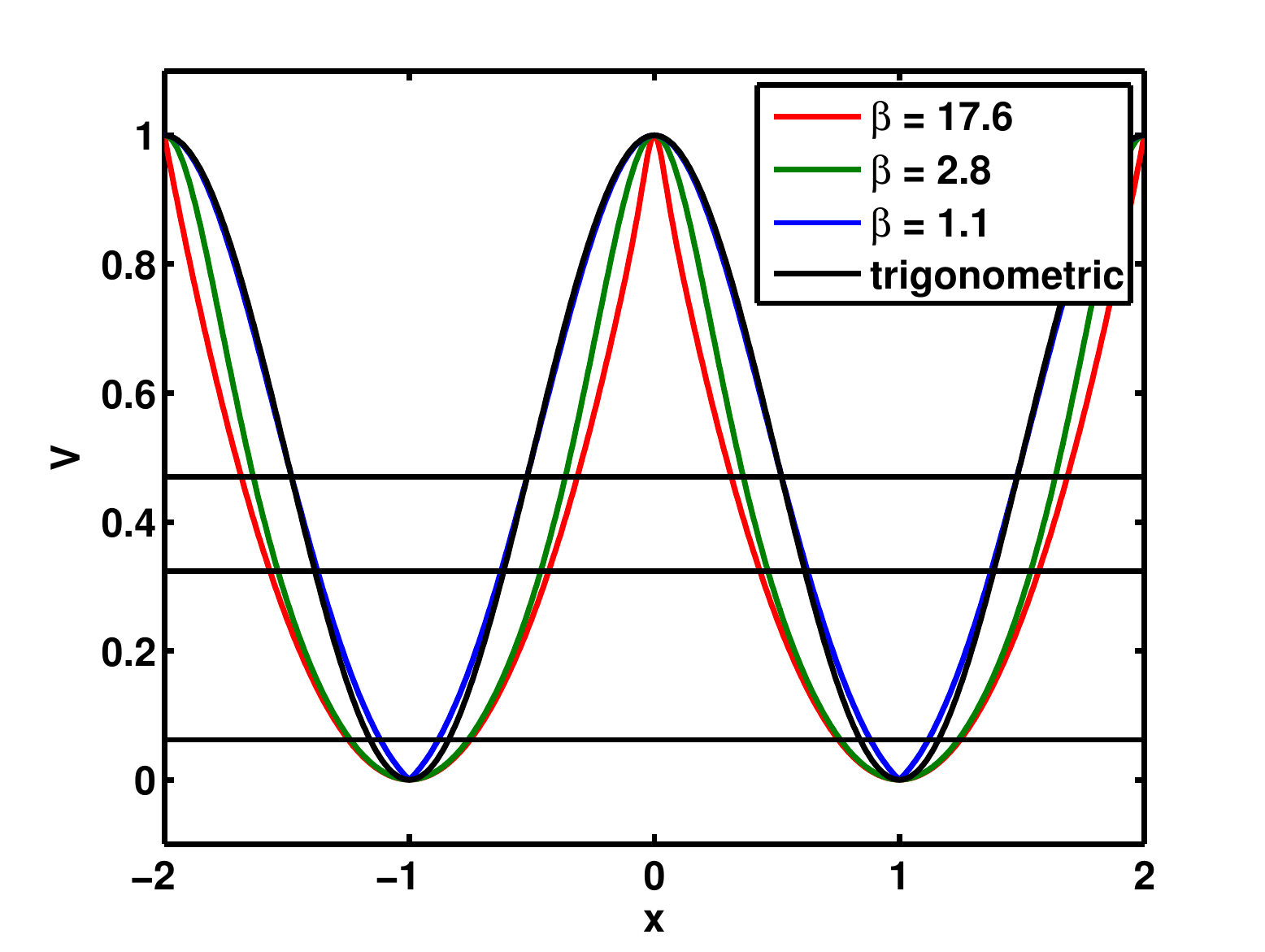}
\caption{Scaled symmetric double well potential $V(x)$ for the cyclic Gaussian model with dimensionless 
action parameter $\beta$ as specified in the legend, eqn. (\ref{eqn:betaphi}). The black curve represents the 
trigonometric potential used in the Mathieu model (\ref{eqn:mamo}).
Horizontal black lines indicate the zero point energies of the Mathieu model for
(from top) $\beta = 1.1, 2.8$ and $17.6$, respectively.}
\label{fig:pot_c}
\end{figure}

The cyclic potentials $V(x)$ for $\beta = 1.1, 2.8$ and $17.6$, are shown in Figure \ref{fig:pot_c}. 
As for the non-cyclic case, the potential for $\beta = 1.1$ is not suitable, and enforcing the periodicity 
conditions leads to a discontinuous first derivative at $x = \pm 1$. However, already the potential for 
$\beta = 5$ has the desired properties, i.e.\ correct position of minima and maxima and (in very
good approximation) continuous derivatives. 

\subsection{Cyclic Mathieu model}

\begin{figure}[htbp]
\centering
\includegraphics[width=08cm]{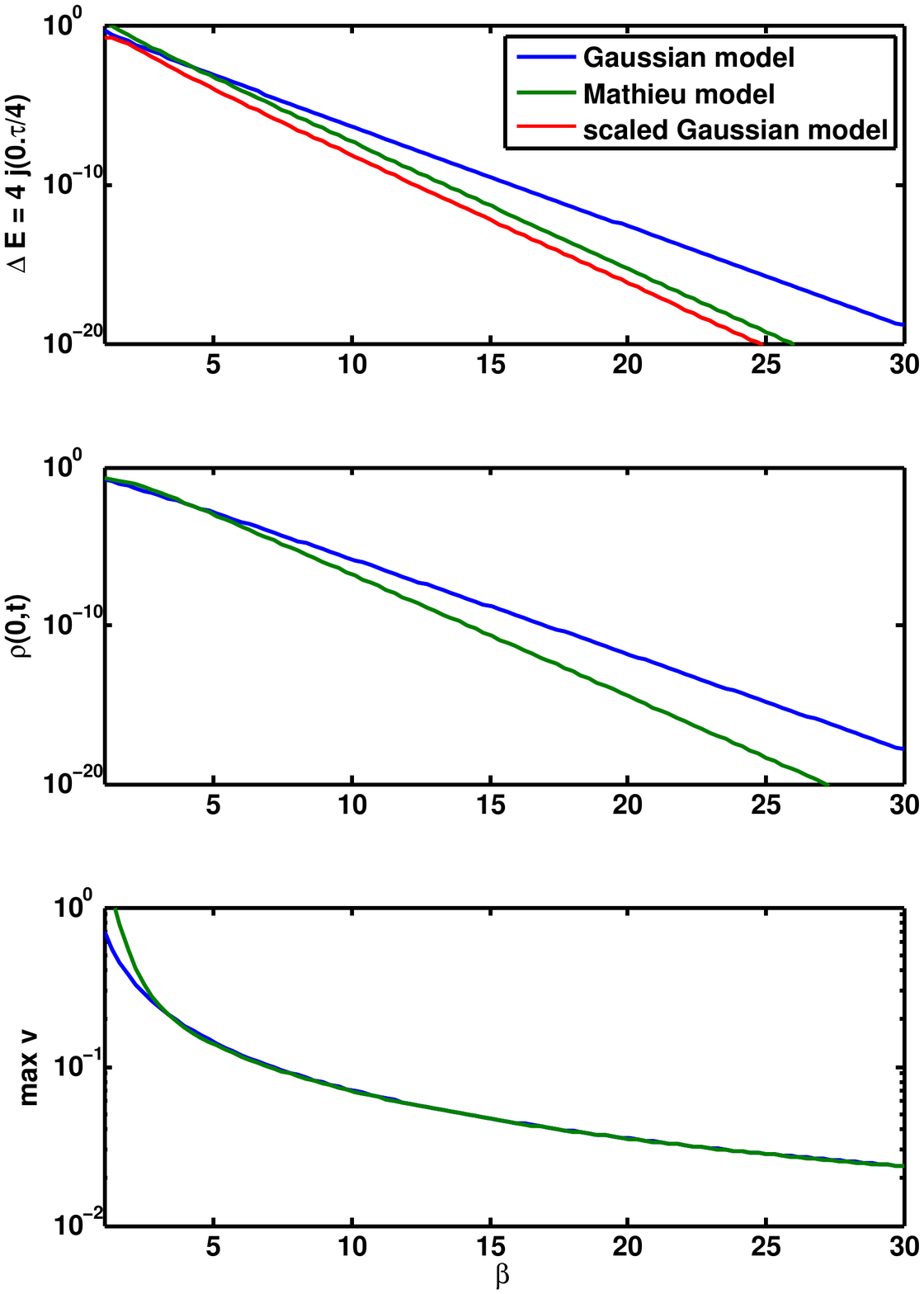}
\caption{Dependence of characteristic quantities on dimensionless action parameter $\beta$, eqns. (\ref{eqn:betax}), (\ref{eqn:betaphi}), for the Gaussian model (\ref{eqn:tunnelsplitscaled}) and for the Mathieu model (\ref{eqn:dem}), as well as for a scaled Gaussian model. Top: tunnel splitting $\Delta E=4j(0,\tau/4)$ for cyclic models ($\Delta E=2j(0,\tau/4)$ for non-cyclic models). Middle: density at the maximum of the barrier (time-independent). Bottom: maximum angular  velocity for tunneling in cyclic models (twice as large for non-cyclic models).
Note that the velocities are given here in terms of $x_0 V_{\rm B}/\hbar$ (non cyclic model) or $\pi V_{\rm B}/(2\hbar)$ (cyclic models). 
However, the corresponding unscaled maximum tunneling velocities increase with the barrier height $V_{\rm B}$ and decrease with mass $m$ or moment of inertia $I$, see eqns. (\ref{eqn:finalmaxvtilde}), (\ref{eqn:finalmaxvdtilde}).}
\label{fig:velocities}
\end{figure}

The cyclic Gaussian model introduced above can be compared with a system characterized by 
a trigonometric potential which obeys the cyclic boundary condition (\ref{eqn:boundarycyclic})
and the previous conditions on the minima and maxima,
\begin{equation}
H = -\frac{1}{2\beta^2} \frac{\partial^2}{\partial x^2} + \frac{1}{2} \cos(\pi x) + \frac{1}{2}, ~~~ x \in [-2,2].
\label{eqn:mamo}
\end{equation}
The corresponding TISE (\ref{eqn:tisescaled}) is equivalent to the Mathieu equation \cite{Mathieu:68a},
\begin{equation}
 \left( \frac{\partial^2 }{\partial \theta^2} - 2 q \cos(2 \theta) + \lambda_n \right) \phi_n(\theta) = 0, ~~~ \theta \in [-\pi,\pi]
\label{eqn:mathieu}
\end{equation}
with $\theta := \frac{\pi}{2} x$, the barrier height
\begin{equation}
 q = 2 \left( \frac{\beta}{\pi} \right)^2 \label{eqn:matq}
\end{equation}
and the eigenvalues
\begin{equation}
 \lambda_n = \frac{4 \beta^2}{\pi^2} (2 E_n - 1). \label{eqn:mate}
\end{equation}
The relation between $\phi_n$ of (\ref{eqn:mathieu}) and $\psi_n$ of (\ref{eqn:tisescaled}) 
is determined by the normalization conditions
\begin{equation}
 \int_{-\pi}^{\pi} |\phi_n(\theta)|^2 d\theta = \int_{-2}^2 |\psi_n(x)|^2 dx \stackrel{!}{=} 1,
\end{equation}
so that 
\begin{equation}
 \psi_n(x) = \sqrt{\frac{\pi}{2}} \phi_n(\theta),
\end{equation}
and the eigenfunctions $\phi_0$ and $\phi_1$ are obtained as the lowest
order Mathieu's cosine elliptic (ce) or sine elliptic (se) functions,
respectively \cite{Mathieu:68a}. Note that these functions are straightforward to obtain
as eigenvalues of a tri-diagonal matrix \cite{Gutierrez:03a}.
In addition, there is a result from asymptotic analysis for the gap between the lowest two eigenvalues $\lambda_0$ and $\lambda_1$ is (see
eq.\ (20.2.31) of \cite{Abramovitz:72a} and paragraph (2.331) of \cite{Meixner:54a})
\begin{equation}
 \lambda_1 - \lambda_0 \approx \sqrt{\frac{2}{\pi}} 2^{5} q^{3/4} \exp(-4 \sqrt{q}).
\end{equation}
Thus, by virtue of (\ref{eqn:mate}), the energy splitting is given in terms of $\beta$ (\ref{eqn:matq}) by 
\begin{equation}
 \Delta E = E_1 - E_0 \approx 8 \frac{2^{1/4}}{\sqrt{\beta}} \exp\left(-\frac{4 \sqrt{2}}{\pi} \beta\right).
 \label{eqn:dem}
\end{equation}
These tunnel splittings are also shown in the upper panel of Figure \ref{fig:velocities}.
For small $\beta$, the tunnel splittings of the Gaussian model and the Mathieu model
agree very well, whereas for larger $\beta$ they start to deviate due to the prefactors
in the exponentials of (\ref{eqn:deltaeweiner}) and (\ref{eqn:dem}), which differ by a factor
of $4/\pi \approx 1.2732$. In order to investigate if this discrepancy is due to 
the shape of the potential, we calculate the barrier integration integrals
\cite{Miller:79a}
\begin{equation}
 S = \int_{-x_1}^{x_1} \sqrt{2 \beta^2 (V(x) - E_0)} dx,
\end{equation}
where $\pm x_1$ are beginning and end of the tunneling region $V(x) > E_0$.
Rescaling the parameter $\beta$ for the Gaussian model with the ratio of $S$ 
for the Mathieu and the Gaussian model yields much improved agreement between
the tunnel splittings for the two models, see the red curve in Figure \ref{fig:velocities}. Hence, 
the difference is mainly caused by the difference of the actions $S$.

\section{Tunneling dynamics}
\label{sec:tudy}

\begin{figure}[htbp]
\centering
\includegraphics[width=08cm]{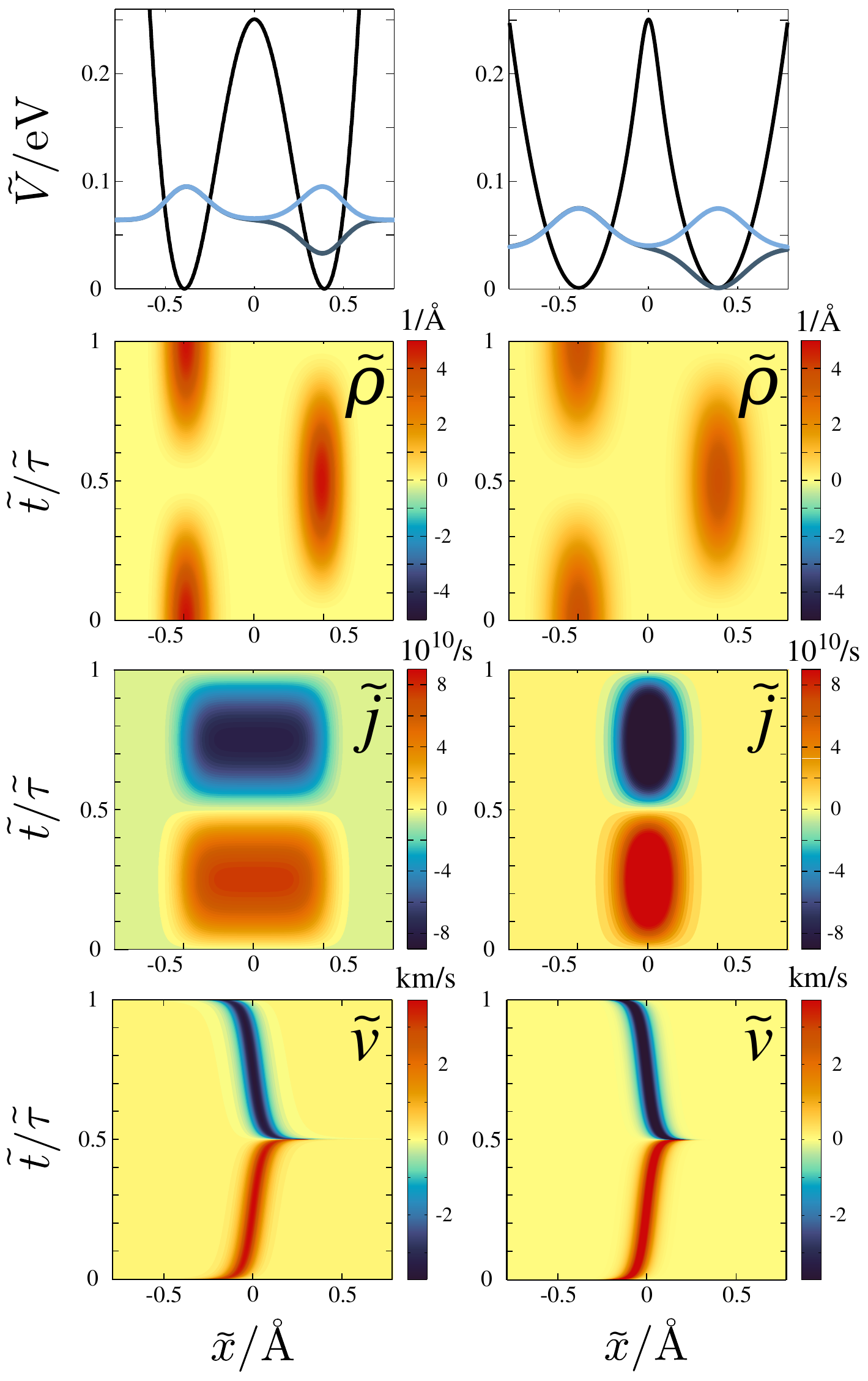}
\caption{Tunneling dynamics for model of ammonia \cite{Letelier:97a} 
(left panel) and non-cyclic Gaussian model with equivalent dimensionless
action parameter $\beta = 4.8$ (right panel). Top to bottom: Unscaled 
potential $\tilde{V}(\tilde{x})$ and wave functions $\tilde{\psi}_0,
\tilde{\psi}_1$ of the lowest tunneling doublet. Spatio-temporal 
representation of density $\tilde \rho$, flux density $\tilde j$, and velocity 
${\tilde v} = {\tilde j}/{\tilde \rho}$, for one tunneling period $\tilde \tau$.}
\label{fig:ammonia}
\end{figure}

Next, we determine the wave function $\psi(x,t)$ which describes 
tunneling in the model system with Hamiltonian $H$ (\ref{eqn:h}). 
For this purpose, we assume that at time $t = 0$, 
the wave function is localized in the left potential well,
\begin{equation}
\psi(x,0) = \psi_{\rm l}(x) = \frac{1}{\sqrt{2}} (\psi_0(x) + \psi_1(x)).
\label{eqn:psil}
\end{equation}
Its time evolution is obtained from (\ref{eqn:tdsescaled}),
subject to the above initial condition, as 
\begin{equation}
\psi(x,t) = \frac{1}{\sqrt{2}} (\psi_0(x) \exp(-i E_0 t) + \psi_1(x) \exp(-i E_1 t)).
\label{eqn:psiyt}
\end{equation}
The corresponding density
\begin{equation}
\rho(x,t) =  |\psi(x,t)|^2 = \frac{1}{2} (|\psi_0|^2 + |\psi_1|^2) + \cos\left(2\pi \frac{t}{\tau}\right) \psi_0 \psi_1
\label{eqn:density}
\end{equation}
oscillates with a tunneling time
\begin{equation} 
\tau = \frac{2\pi}{\Delta E}.
\label{eqn:tau}
\end{equation} 
For the case of $\beta \gg 1$ considered here, the tunneling splittings 
(\ref{eqn:deltaeweiner}) become very small, and the tunneling
time $\tau$ becomes very long, see Table \ref{tab:tablee}.
The density tunnels periodically from the left potential well,
\begin{equation}
\rho(x,t_{\rm l}) = \rho_{\rm l}(x) = |\psi_{\rm l}(x)|^2 
\label{eqn:rhol}
\end{equation}
at times $t_{\rm l} = 0, \tau,2 \tau, \dots$, to the right one,
\begin{equation}
\rho(x,t_{\rm r}) = \rho_{\rm r}(x) = |\psi_{\rm r}(x)|^2 
\label{eqn:rhor}
\end{equation}
at times $t_{\rm r} = \tau/2, 3\tau/2, 5\tau/2, \dots$,
whereas it is symmetrically delocalized,
\begin{equation}
\rho(x,t_d) = \frac{1}{2} (\psi_{\rm l}(x)^2 + \psi_{\rm r}(x)^2 )
        = \frac{1}{2} (\psi_0(x)^2 + \psi_1(x)^2) 
\label{eqn:rhodelocalized}
\end{equation}
at intermediate times $t_{\rm d} =\tau/4, 3\tau/4, 5\tau/4, \dots$. 
Figure \ref{fig:ammonia} shows an example of this dynamics for $\beta = 4.8$.
Of special interest for tunneling dynamics is the value of the 
density at the maximum of the barrier, given by 
$\rho(0,t) = |\psi_0(0)|^2/2$, independent of time. 
For the Gaussian model (\ref{eqn:psi0}), (\ref{eqn:psilr}), (\ref{eqn:b1}), (\ref{eqn:deltaeweiner}
)
we find
\begin{eqnarray}
\rho(0,t) &=& 2^{1/4} \sqrt{\frac{\beta}{\pi}} \exp(-\sqrt{2}\beta) \nonumber \\
          &=& \frac{\beta}{\sqrt{2}} \Delta E,
\label{eqn:rhomin}
\end{eqnarray}
which is compared with the numerical values obtained for the Mathieu model
in the middle panel of Figure \ref{fig:velocities}, where the latter one decays 
faster with increasing $\beta$. This behavior can be explained by the 
shape of the potentials at the barrier, see also Figures \ref{fig:pot_nc},
\ref{fig:pot_c}.

To further characterize the tunneling dynamics, we calculate the corresponding probability
flux density \cite{Schroedinger:26a}. For our scaled model systems, it is defined as
\begin{equation}
j(x,t) = \frac{1}{\beta^2} \operatorname{Im}\left(\psi^*(x,t) \frac{\partial}{\partial x} \psi(x,t)  \right).     
\label{eqn:j}
\end{equation}
Near the barrier at $x = 0$, however, this is amenable to considerable 
round-off errors, especially when $\beta$ becomes large. Hence, with the help 
of the continuity equation this expression is rewritten as \cite{Bredtmann:13a,Grohmann:13a}
\begin{equation}
j(x,t) = \Delta E \sin\left(2\pi \frac{t}{\tau}\right) \int_{-\infty}^x \psi_0(x') \psi_1(x') dx'.
\label{eqn:jnew}
\end{equation}
For our cyclic model, the lower boundary of 
the integration has to be replaced by $x = -1$, where 
$j = 0$ by symmetry. Note that the sign change of the integrand at $x = 0$ implies
that $j$ reaches its maximum value w.r.t.\ $x$ at that point. Also, because for 
large values of $\beta$, the magnitude of the wave functions $\psi_0$ and $\psi_1$ 
are approximately equal for $x < 0$, 
\begin{equation}
 j(0,t) \approx \frac{\Delta E}{2} \sin\left(2\pi \frac{t}{\tau}\right)
 \label{eqn:absj}
\end{equation}
for non-cyclic systems, which is illustrated in third row of Figure \ref{fig:ammonia}. 
For cyclic systems, this flux density is half as large, because density moves equally 
to the left and to the right \cite{Grohmann:13a}.
Note that the $\beta$-dependence of $\Delta E$ in our Gaussian models is given by (\ref{eqn:deltaeweiner}).

From the probability density and its flux density, the velocity during tunneling can be 
calculated in analogy to the classical definition of the flux density
\begin{equation}
v(x,t) = \frac{j(x,t)}{\rho(x,t)}   
\label{eqn:v}
\end{equation} 
which is used in our subsequent calculations of tunneling velocities {\cite{Winful:06a,Tannor:07,Bohm:52a,Wyatt:05a}}. 
The (linear or angular) tunneling velocities achieve their maxima at the potential barriers ($x=0$) at one quarter of the tunneling time, $t=\tau/4$ (with periodic recurrences at $t = 5 \tau /4, 9\tau/4$ etc). 
The maximum tunneling velocities are, therefore,
\begin{equation}
\max v(x,t) = \frac{j(0,\tau/4)}{\rho(0,\tau/4)}
\label{eqn:maxv}.
\end{equation}

For the Gaussian model with initial condition (\ref{eqn:psil}), the
maximum tunneling velocity depending on the parameter $\beta$ shall 
now be calculated. At $x = 0$ and $t = \tau/4$ the probability flux 
density (\ref{eqn:jnew}) for the non-cyclic system is
\begin{equation}
j(0,\tau/4) = \frac{\Delta E}{2},
\label{eqn:jmax}
\end{equation}
and half of this expression for the cyclic system.
Again, the reader is reminded that the $\beta$-dependence of $\Delta E$ for our Gaussian models is given in (\ref{eqn:dem}).
Together with the density (\ref{eqn:rhomin}), we obtain our final results, i.e.\
the maximum tunneling velocity
\begin{equation}
\max v(x,t) = \frac{\sqrt{2}}{\beta}                                         
\label{eqn:finalmaxv}
\end{equation}
for the tunneling in the non-cyclic double well system in terms of the units
$x_0 V_{\rm B}/\hbar$, as well as the maximum angular tunneling velocity
\begin{equation}
\max v(x,t) = \frac{1}{2}\frac{\sqrt{2}}{\beta}                                        
\label{eqn:finalmaxvcyclic}
\end{equation}
for the cyclic one, in terms of the units $V_{\rm B}/\hbar$. 
The bottom panel of Figure \ref{fig:velocities} shows a comparison between the maximum velocities for the 
Gaussian model and the Mathieu model. 
It can be seen that the agreement between the two models for large $\beta$ is very good. 
Thus, the shape of the potential affects the density and via the density also the flux density at the maximum of the potential, but not the velocity.

Before closing this Section, let us remark that, rewardingly, the expression (\ref{eqn:v}) for the velocity can also be used in order to derive an estimate $\tau'$ of the tunneling time $\tau$. 
For this purpose, {let us consider first, for reference,} the scenario when the flux density (\ref{eqn:jnew}) achieves its maximum value, i.~e., at the reference time $t_r=\tau/4=\pi/(2 \Delta E)$, and we evaluate the approximate reference tunneling time $\tau'_r$ as twice the time which the particle needs to tunnel from the minimum of the left potential well to the right one, with velocity $v(x,\tau/4)$. Accordingly,
\begin{eqnarray}
\tau'_r &=& 2 \int_{-1}^{+1} dx \frac{1}{v(x,\tau/4)} \nonumber \\
      &=& 2 \int_{-1}^{+1} dx \frac{\rho(x,\tau/4)}{j(x,\tau/4))}\nonumber \\
      &=&   \int_{-1}^{+1} dx \frac{\psi_0^2(x) + \psi_1^2(x)}{\Delta E \int_{-\infty}^x dx' \psi_0(x')\psi_1(x')}\nonumber \\
      &=& 2 \int_{-1}^{ 0} dx \frac{\psi_0^2(x) + \psi_1^2(x)}{\Delta E \int_{-\infty}^x dx' \psi_0(x')\psi_1(x')}
\end{eqnarray}
where we have used (\ref{eqn:v}) and (\ref{eqn:rhodelocalized}), (\ref{eqn:jnew}) in the second and third step, respectively, and where we have exploited the symmetries of the wavefunctions $\psi_0(x)$ and $\psi_1(x)$ in the last step. 
Since their shapes are rather similar in the left potential well, 
\begin{eqnarray}
\tau'_r &\approx& 4 \int_{-1}^{0} dx \frac{\psi_0^2(x)}{\Delta E \int_{-\infty}^x dx' \psi_0^2(x')} \nonumber \\
      &   =   &  \frac{4}{\Delta E} \int_{-1}^{ 0} dx \frac{u'(x)}{u(x)} \nonumber \\
      &   =   &  \frac{4}{\Delta E} (\ln u(0)-\ln u(-1)) \nonumber \\
      &   =   &  \frac{4}{\Delta E} (\ln(1/2) -\ln(1/4) ) \nonumber \\
      &   =   &  \frac{4 \ln 2}{\Delta E} \approx \frac{2.77}{\Delta E}
      \label{eqn:taudash}
\end{eqnarray}
where $u(x)=\int_{-\infty}^x dx' \psi_0^2(x')$ such that $u(0)=1/2$ and $u(-1)=1/4$ in the deep tunneling regime. 
This estimate of the reference tunneling time $\tau'_r$ is close to $\tau$ as given in eqn. (\ref{eqn:tau}).
Now the estimate $\tau'_r$ at reference time $t_r$ corresponds to the maximum velocity, that means $\tau'_r$ is a lower limit of $\tau'$. A better "mean" estimate $\tau'$ is obtained by averaging the corresponding expressions over all times $0\le t \le \tau/2$, thus
\begin{equation}
\tau' = \int_0^{\tau/2} dt \,2 \int_{-1}^1 dx \frac{1}{v(x,t)\tau/2}
\end{equation}
Of course, the result for $\tau'$ will be somewhat larger than the reference value $\tau'_r$ obtained for the maximum velocity. 
Gratifyingly, however, the s-shaped contour plot of $v(x,t)$ during tunneling from the left to the right potential wells, i.~e., during $0 \le t \le \tau/2$, as documented in {Figure~\ref{fig:ammonia}} shows that the profile of $v(x,t)$ versus $x$ is robust for almost all times, except close to the beginning and to the end of the tunneling processes. 
This means that the estimate $\tau'_r$ derived for $t_r = \tau/4$ actually serves as a rather good estimate for almost all other times. 
Averaging over all times will, therefore, yield the mean value $\tau'$ which is just slightly larger than $\tau'_r$, say by a {dilatation factor of about two}. 
This may well account for the difference of the factor $4 \ln 2$ in eqn. (\ref{eqn:taudash}) versus $2 \pi$ in eqn. (\ref{eqn:tau}). 
Finally, perfect agreement could be obtained if one evaluates the time for passing from $x=-a$ slightly left of the minimum of the left potential well to $x=+a$, slightly right of the minimum of the right potential well. 
This suggests that one may analyze the total tunneling time in terms of the time intervals which are needed in order to tunnel through certain domains of the barrier. 
In particular, we are interested in the time $t_B$ for tunneling through the top of the potential barrier.
In other words, this is the time which the particle needs in order to tunnel through the "most difficult part" of the potential $V(x)$, i.~e., the domain just below the barrier, say from $x=-1/10$ to $x=+1/10$. 
Here, $v(x,\tau/4)$ is close to the maximum velocity $\max v$, hence 
\begin{equation}
t_B = \int_{-1/10}^{+1/10} \frac{dx}{\max v(x,t)} = \frac{1}{5 \max v(x,t)}. 
\label{eqn:t_B}
\end{equation}
\section{Applications, Discussion, and Conclusions }
\label{sec:discuss}
In order to understand our findings, we convert eqns.\ (\ref{eqn:finalmaxv}), (\ref{eqn:finalmaxvcyclic}) 
back to the unscaled original systems. This yields
\begin{equation}
\max \tilde{v} = \sqrt{\frac{2 V_{\rm B}}{m}}
\label{eqn:finalmaxvtilde}
\end{equation}
for the non-cyclic and
\begin{equation}
\max \tilde{v} = \frac{1}{2} \sqrt{\frac{2 V_{\rm B}}{I}},  
\label{eqn:finalmaxvdtilde}
\end{equation}
for the cyclic Gaussian model, respectively. 
{Expression (\ref{eqn:finalmaxvtilde}) reminds of the maximum absolute value of the imaginary semiclassical Wentzel-Kramers-Brillouin-Jeffreys (WKBJ) velocity
\begin{eqnarray}
 \tilde{v}_{\rm WKBJ,max} & = & \max \left| i \sqrt{2 m (V(x) - E)} \right| \nonumber \\
& = & \sqrt{2 m (V(0) - E)}
\end{eqnarray}
in the deep tunneling domain where $E \approx E_0$, thus $V(0) - E \approx V_{\rm B}$ \cite{Makri:89,Garg:99,Razavy:03,Schwabl:07}.}
Various applications are documented in Table \ref{tab:tablee}. 
Remarkably, the maximum tunneling 
velocity (\ref{eqn:finalmaxvtilde}) in the deep tunneling {regime
depends} on the barrier height $V_{\rm B}$ and the mass $m$ or moment of inertia 
$I$, but not on the potential width $x_0$.
The maximum tunneling velocity increases with increasing barrier height and decreases with increasing mass
or moment of inertia, despite of the increasing tunneling times. 
At first glance, the maximum velocities (\ref{eqn:finalmaxvtilde}) (or angular velocities (\ref{eqn:finalmaxvdtilde})) appear to be unbounded, possibly pointing to values which might approach, or even exceed, the speed of light. 
However, this premature conjecture is wrong for two reasons. 
First, the relativistic masses (or the corresponding moments of inertia) of particles approach infinity as their velocities approach the speed of light, implying upper limits of the maximum velocities (or the corresponding angular velocities) below the speed of light, cf. the right hand side of eqn. (\ref{eqn:finalmaxvtilde}) or (\ref{eqn:finalmaxvdtilde}). 
Second, the parameters of typical chemical systems compiled in Table~\ref{tab:tablee} imply that the maximum  velocities are in the order of a few km/s (or THz for angular dynamics) for typical chemical systems. 
Even for the extreme case of an electron tunneling through a barrier of one Hartree (\unit[27.2]{eV}), the maximum velocity is still only one percent of the speed of light which -- a posteriori -- justifies our non-relativistic approach and safely excludes the superluminality. 

It may appear a bit paradoxical that the maximum tunneling velocity increases with $V_{\rm B}$, 
although the tunneling times $\tau$ also increase with $V_{\rm B}$. 
One may rationalize this intuitively by saying that the higher $V_{\rm B}$, the more 
difficult is the tunneling through a region of extremely low density, 
which has to be compensated by higher velocity.
For a more quantitative analysis,we consider the times for tunneling through the top of the potential barrier,
{say from $-x_0/10$ to $x_0/10$}, eqn. (\ref{eqn:t_B}). In SI units, 
\begin{equation}
{\tilde t}_B \approx \frac{x_0}{5 \max \tilde{v}}
\label{eqn:tB1}
\end{equation} 
for non-cyclic systems. Typical values from Table \ref{tab:tablee} correspond to \unit[2-4]{fs}.
For cyclic systems,
\begin{equation}
{\tilde t}_B \approx \frac{\pi}{10 \max \tilde{v}},
\label{eqn:tB2}
\end{equation}
and typical values from Table \ref{tab:tablee} are in the domain of \unit[2-40]{fs}. 
In both cases, these times are several orders of magnitude smaller than the tunneling time $\tilde \tau$. 
{The ratio
\begin{equation}
 \frac{\tilde{t}_{\rm B}}{\tilde{\tau}/2} = 0.1 \times \frac{2 x_0}{\tilde{\tau} \max \tilde{v}}
\end{equation}
is thus exceedingly small, similar to the ratio of the time-averaged tunneling velocity 
$\operatorname{avg} \tilde{v} = 2 x_0 / (\tilde{\tau}/2)$ versus the maximum tunneling velocity,
\begin{equation}
 \frac{\operatorname{avg} \tilde{v}}{\max \tilde{v}} = \frac{2 x_0}{\tilde{\tau} \max \tilde{v}},
\end{equation}
see Table \ref{tab:tablee}.}
In a possible experiment monitoring the velocities of the particles under the barrier, one would have to wait on the average as long as the tunneling time until the density has moved from one well to the other.
In fact, for some of the systems with rather high potential barrier $V_B$, the tunneling times are so long that in practice one would have to wait "forever" -- for these systems, tunneling is irrelevant. 
Nevertheless, we have included these examples in Table \ref{tab:tablee} in order to demonstrate that even for these extreme scenarios, the maximum tunneling velocities are well below the {speed of light. 
However,} the event of tunneling through the domain under the top of the barrier actually takes much faster -- it is in the fs time domain for molecular processes involving nuclear dynamics.
The situation reminds of other quantum processes in physics and chemistry which involve two very different time scales, e.g. the mean radiative life times of molecules which are typically in the ns or even longer time domains, compared with the time for the transition from the electronic excited to the ground state during the emission of the photon, which may occur within atto- or perhaps femtoseconds {\cite{Haroche:13}}. 

The present results for coherent tunneling in simple models with double well potentials along a single coordinate $\tilde x$ should stimulate extended investigations {in terms of complementary approaches, e.g.\ quantum trajectory or semiclassical methods \cite{Takatsuka:99,Bittner:00,Ushiyama:05,Poirier:08}}, of models in full dimensionality, as well as the development of experimental methods for monitoring the predicted tunneling velocities. 
The results may also serve as a reference for tunneling velocities during incoherent tunneling in systems with symmetric double well potentials coupled to an environment.

\section*{Acknowledgments}
We are grateful for financial support of our project by the Deutsche Forschungsgemeinschaft under Grants No. Ma 515/23-3, Ma 515/25-1, the 973 Program of China under Grant No. 2012CB921603, and the National Natural Science Foundation of China under Grants Nos. 11004125 and 10934004. 
{We thank Prof. Bretislav Friedrich (Fritz Haber Institut Berlin) for valuable hints to the literature.}
J.M. would also like to express thanks to Sir Michael Berry {(University of Bristol)} for a stimulating discussion remark after a lecture on intramolecular flux densities in symmetric double well potentials (Universit{\"a}t Kaiserslautern, April 2013), suggesting that the results should imply rather large tunneling velocities. 
J.M. also thanks Professor Beate Paulus (Freie Universit{\"a}t Berlin) for pointing to limitations of the maximum tunneling velocities below the speed of light due to relativistic increases of the masses of ultrafast particles. 
We are grateful to Professor Omar Deeb (AlQuds University), Professor Leticia Gonz{\'a}lez, Dr. Markus Oppel, Mrs. Rana Obaid (Universit{\"a}t Wien) and Profesor Shmuel Zilberg (Hebrew University Jerusalem) for cooperation on torsional quantum dynamics. 

\section*{References}
\bibliography{Tunneling}{}
\bibliographystyle{model1-num-names}

\clearpage
\begin{sidewaystable}[htbp]
 \centering
 {\small
 \begin{tabular}{l|cccccccc}
system                                                                                      & $V_{\rm B}$            & $m$ or $I$               & $x_0$            & $\beta$ & $\max \tilde{v}$  & {$\operatorname{avg} \tilde{v}$} & $\tilde\tau$       & ${\tilde t}_{\rm B}$ \\
\hline
ammonia tunneling \cite{Letelier:97a}                                                       & \unit[0.250]{eV}       & \unit[2.49]{u}           & \unit[0.39]{\AA} &  4.8    & \unit[4.4]{km/s}  &  {\unit[9.8]{m/s}}               & \unit[15.9]{ps}    & \unit[1.8]{fs}       \\
semibullvalene Cope rearrangement \cite{Zhang:10a,Andrae:11a,Bredtmann:11a,Bredtmann:13a}   & \unit[0.355]{eV}       & \unit[3.25]{u}           & \unit[0.96]{\AA} & 15.9    & \unit[4.6]{km/s}  &  {\unit[2.5]{$\mu$m/s}}          & \unit[153]{$\mu$s} & \unit[4.2]{fs}       \\
para- $\leftrightarrow$ ortho-fulvene \cite{Grohmann:07a}                                   & \unit[1.94]{eV}        & \unit[1.7]{u\! \AA$^2$}  & $\pi/2$          & 44.1    & \unit[74]{THz}    &  {\unit[66.8]{pHz}}              & \unit[9.4]{Ts}     & \unit[4.2]{fs}       \\
methaniminium cis-trans isomerization  \cite{Barbatti:08a}                                  & \unit[3.65]{eV}        & \unit[1.21]{u\! \AA$^2$} & $\pi/2$          & 51.1    & \unit[121]{THz}   &  {\unit[64.1]{fHz}}              & \unit[98]{Ps}      & \unit[2.6]{fs}       \\
quinodimethane derivative torsion A  \cite{Belz:13a}                                        & \unit[2.68]{eV}        & \unit[39.5]{u\! \AA$^2$} & $\pi/2$          & 250     & \unit[18]{THz}    &  {``0''}                         & ``$\infty$''       & \unit[17.4]{fs}      \\
quinodimethane derivative torsion C  \cite{Belz:13a}                                        & \unit[1.94]{eV}        & \unit[42.2]{u\! \AA$^2$} & $\pi/2$          & 220     & \unit[15]{THz}    &  {``0''}                         & ``$\infty$''       & \unit[21.1]{fs}      \\
nitrogen alignment $10^{14}$ TW/cm$^2$ \cite{Owschimikow:09a}                               & \unit[0.12]{eV}        & \unit[8.47]{u\! \AA$^2$} & $\pi/2$          & 24.1    & \unit[8.1]{THz}   &  {\unit[0.11]{Hz}}               & \unit[59]{s}       & \unit[38.6]{fs}      \\
for comparison: electron (non-periodic)                                                     & \unit[1]{E$_{\rm h}$}  & \unit[1]{m$_{\rm e}$}    & \unit[1]{a$_0$}  & 1       & \unit[3094]{km/s} &  {\unit[643]{km/s}}              & \unit[329]{as}     & \unit[3.4]{as}
 \end{tabular}}
 \caption{Barrier heights $V_{\rm B}$, masses $m$ / moments of inertia $I$, 
 width parameter $x_0$, and dimensionless action parameter $\beta$ for example systems characterized by tunneling in a symmetric double well potential, eqns (\ref{eqn:betax}),(\ref{eqn:betaphi}). The last {four} columns give our results for the maximum tunneling velocity $\max \tilde v$, {the time-averaged tunneling velocity $\operatorname{avg} \tilde{v}$}, the tunneling time $\tilde \tau$ and the time ${\tilde t}_{\rm B}$ needed for tunneling through the top of the potential barrier, as defined in equations (\ref{eqn:finalmaxvtilde},\ref{eqn:finalmaxvdtilde}), (\ref{eqn:tau}), and (\ref{eqn:tB1},\ref{eqn:tB2}), respectively.}
 \label{tab:tablee}
\end{sidewaystable}

\clearpage

\clearpage

\clearpage

\clearpage

\end{document}